\documentclass[%showpacs,
showkeys,12pt,
preprint,preprintnumbers,nofootinbib,
groupedaddress,superscriptaddress,amsmath,amssymb]{revtex4}
\usepackage{graphicx}% Include figure files
\usepackage{dcolumn}% Align table columns on decimal point
\usepackage{bm}% bold math
\usepackage{amssymb}
\usepackage{amsmath}
\usepackage{epsfig}    
\usepackage{color}
\usepackage{slashed}
\usepackage{hhline}
\def\be{\begin{equation}}
\def\ee{\end{equation}}
\newcommand{\bea}{\begin{eqnarray}}
\newcommand{\eea}{\end{eqnarray}}
\newcommand{\nn}{\nonumber}

\numberwithin{equation}{section}

\begin{document}

%%%%%%%%%
\title{
A Three-Loop Neutrino Model with Global $U(1)$ Symmetry
%Three Loop Neutrino Model and Dark Matters with Global $U(1)'$ Symmetry
}
\preprint{KIAS-P14079}

\author{Hisaki Hatanaka}
\email{hatanaka@kias.re.kr}
\affiliation{School of Physics, KIAS, Seoul 130-722, Republic of Korea}

\author{Kenji Nishiwaki}
\email{nishiken@kias.re.kr}
\affiliation{School of Physics, KIAS, Seoul 130-722, Republic of Korea}

\author{Hiroshi Okada}
\email{hokada@kias.re.kr}
\affiliation{School of Physics, KIAS, Seoul 130-722, Republic of Korea}

\author{Yuta Orikasa}
\email{orikasa@kias.re.kr}
\affiliation{School of Physics, KIAS, Seoul 130-722, Republic of Korea}
\affiliation{Department of Physics and Astronomy, Seoul National University, Seoul 151-742, Republic of Korea}

\date{\today}

\begin{abstract}

We study a three-loop induced neutrino model with a global $U(1)$ symmetry at TeV scale, in which we naturally accommodate a bosonic dark matter candidate. We discuss the allowed regions of masses and quartic couplings for charged scalar bosons as well as the dark matter mass on the analogy of the original Zee-Babu model, and show the difference between them.
We also discuss {that} the possibility of the collider searches in a future like-sign electron liner collider could be promising.

\end{abstract}
\maketitle
\newpage

\section{Introduction}

Even after the discovery of the Higgs boson, the large Yukawa hierarchy required {by} the observed values of the fermion masses remains to be one of the unnatural issues %a problem
in the Standard Model (SM). The situations get to be more serious in the neutrino sector since their corresponding values are {sub-eV}, which means that we have to realize at least $\mathcal{O}(10^{11})$-magnitude hierarchy by hand when we adapt the Dirac-type mass terms for explanation.
An elegant way for alleviating the unnaturalness is making the situation that the neutrino masses are loop-induced as initiated by A.~Zee at one-loop level in Ref.~\cite{Zee}.

In such a setting, loop factors naturally reduce their mass values and we can explain the minuscule neutrino masses with less fine-tuned Yukawa couplings.
This mechanism is fascinating and lots of works have been done in this direction~\cite{Zee}--\cite{Jin:2015cla}.
As a naive expectation, higher-loop generated neutrino masses would be preferable because much more improvement could be expected due to a large amount of loop factors.
Several three-loop models have been proposed already, e.g., in Refs.~\cite{Krauss:2002px,Aoki:2008av,Gustafsson,Kajiyama:2013lja,Chen:2014ska}.
In higher-loop models, a dark matter (DM) candidate tends to propagate inside the loop, whose stability is naturally ensured by symmetries for prohibiting lower-level neutrino masses.
Also, when a continuous global symmetry is used in such a model, we would predict a Nambu-Goldstone boson (NGB).
This kind of {particles could} play a significant role in an early stage of the {Universe}~\cite{Weinberg:2013kea}.

In this paper, we propose a model as a simple extension of the Zee-Babu model~\cite{zee-babu} with two-loop induced neutrino mass terms, by adding an additional singly-charged gauge singlet scalar and DM to the original one, {where the radiative neutrino mass terms turn out to appear at the three-loop level}.
Note that a doubly-charged scalar ($k^{\pm\pm}$) and a {singly-charged} singlet scalar ($h^\pm$) are introduced in the Zee-Babu model~\cite{zee-babu}.
Our model overcomes a shortcoming in the Zee-Babu model of the absence of DM candidate.
{On the other hand, the structure of the internal loops within the radiative neutrino masses gets to be morphed.
Therefore, expected mass ranges of the charged particles are affected from the original ones.}

This paper is organized as follows.
In Sec.~\ref{sec:mainbody}, we explain the construction of our model and analyze the system with declaring brief prospects in collider-related issues.
We summarize and conclude in Sec.~\ref{sec:conclusions}.

%This paper is organized as follows.
%In Sec.~II, we show our model building including Higgs potential, neutrino masses, and muon anomalous magnetic moment.
%In Sec.~III, we analyze DM properties including relic density and the direct detection with multicomponent scenario. We summarize and conclude in Sec.~VI.

\begin{widetext}
\begin{center} 
\begin{table}[tbc]
%\begin{tiny}
\begin{tabular}{|c||c|c||c|c|c|c|c|c|}\hline\hline  
&\multicolumn{2}{c||}{Lepton {fields}} & \multicolumn{6}{c|}{Scalar {fields}} \\\hline
& ~$L_L$~ & ~$e_R^{}$ & ~$\Phi$~ &
 ~$\Sigma_0$~ & ~$h^+_1$~  & ~$h^{+}_2$~ & ~$k^{++}$ & ~$\chi_0$ \\\hline 
$SU(2)_L$ & $\bm{2}$ & $\bm{1}$&  $\bm{2}$&$\bm{1}$ & $\bm{1}$ &$\bm{1}$  &$\bm{1}$  &$\bm{1}$\\\hline 
$U(1)_Y$ & $-1/2$ & $-1$ & $1/2$ & $0$  & $1$  & $1$ & $2$  & $0$  \\\hline
$U(1)$ & $-x$ & $-x$ & $0$ & $x$  & $2x$ & $x$ & $2x$  & $-x$ \\\hline
%%%
$\mathbb{Z}_2$ & $+$ & $+$ & $+$ & $+$  & $+$  & $-$ & $+$  & $-$  \\\hline\hline
\end{tabular}
\caption{Contents of lepton and scalar fields
and their charge assignment under $SU(2)_L\times U(1)_Y\times U(1)\times\mathbb{Z}_2$, where $x\neq 0$.}
\label{tab:1}
% \end{tiny}
\end{table}
\end{center}
\end{widetext}

%In appendices, we show the explicit Higgs potential and ...

%\newpage

%%%%%%%%%%%%%%%%%%%%%%%%%%%%%%%%%%%%%
%\section{The Model}
%\subsection{Model setup}

\section{Discussions on our model\label{sec:mainbody}}

\subsection{Model setup}

We discuss a three-loop induced radiative neutrino model. 
The particle contents and their charges are shown in Tab.~\ref{tab:1}. 
We add new bosons, which are, two {$SU(2)_L$} singlet neutral bosons ($\Sigma_0$, $\chi_0$), two
singly-charged  singlet scalars ($h^+_1, h^{+}_2$), and one {$SU(2)_L$} singlet {doubly-charged} boson $k^{++}$ to the SM.
%%% 
We assume that  only the SM-like Higgs $\Phi$ and $\Sigma_0$ have vacuum
expectation values (VEVs), which are symbolized by $\langle\Phi\rangle\equiv v/\sqrt{2}$  and $\langle\Sigma_0\rangle\equiv v'/\sqrt{2}$, respectively. 
%We also introduce a $U(1)$ symmetry, under which $\Phi$ does not have the charge in order not to couple to the goldstone boson (GB)~\cite{Baek:2014awa}.
$x\,(\neq0)$ is an arbitrary number of the charge of the global $U(1)$ symmetry~\footnote{This symmetry cannot be gauged because its anomaly cannot be {canceled}.}, and the assignments can realize our neutrino model at the three-loop level~\footnote{{Notice here that one can realize our model by assigning zero global $U(1)$ charge to $\chi_0$ instead of $-x$, where $\chi_0$ can be still a (real) DM candidate due to the $\mathbb{Z}_2$ symmetry. Then the following two relevant terms $h^+_1 h^-_2 \chi_0$ and $ (\Sigma_0)^2 (\chi_0)^2$ are respectively replaced by $\Sigma_0^* h^+_1 h^-_2 \chi_0$ and $ (\chi_0)^2$. Such a mechanism has been done by the authors in Ref.~\cite{Kanemura:2012rj}. We would like to thank our referee to draw our attention.}}.
%Otherwise the $\mathbb{Z}_2$ symmetry which guarantees DM stability is spontaneously broken. 
Notice here that one can identify the global $B-L$ symmetry in case $x=1$.
{The NGB in $\Sigma_0$ due to breaking of the $U(1)$ global symmetry can also evade experimental searches or constraints due to its very weak interactions with matter fields as can be seen in \cite{majoron}, when this symmetry is identified as the $L$ symmetry.}
The {$\mathbb{Z}_2$} symmetry assures the stability of DM $\chi_0$.

The relevant Lagrangian for Yukawa sector, mass {terms}, and scalar potential
under these symmetries are given by
\begin{align}
-\mathcal{L}_{Y}
&=
y_\ell \bar L_L \Phi e_R  + y_{L} \bar L^c_L L_L h^+_1   + y_{R} \bar e^c_R e_R k^{++}  +\rm{h.c.}, \\ 
%%%
\mathcal{V}
&= 
 m_\Phi^2 |\Phi|^2 + m_{\Sigma}^2 |\Sigma_0|^2 + m_{h_1}^2 |h^+_1|^2  + m_{h_2}^2 |h_2^{+}|^2   + m_{k}^2 |k^{++}|^2  +  m_{\chi_0}^2 |\chi_0|^2
 \nn\\
&+ \Bigl[
 \mu_{12} h^+_1 h^-_2 \chi_0 +  \mu_{22} h^+_2 h^+_2 k^{--}   + \lambda_0(\Sigma_0)^2 (\chi_0)^2 + {\rm h.c.}\Bigr]
 \nn\\
&
  +\lambda_\Phi |\Phi|^{4} 
 %%%
  + \lambda_{\Phi\Sigma} |\Phi|^2|\Sigma_0|^2 
%%%
 +\lambda_{\Phi h_1}  |\Phi|^2|h^+_1|^2  %+\lambda_6'  |\Phi|^2|\chi^+_2|^2
 \nn\\
& +\lambda_{\Phi h_2}  |\Phi|^2|h^+_2|^2 +\lambda_{\Phi k}  |\Phi|^2|k^{++}|^2  +\lambda_{\Phi \chi_0}  |\Phi|^2|\chi_0|^2
 \nn\\
&
  + \lambda_{\Sigma} |\Sigma_0|^{4} +
 \lambda_{\Sigma h_1}  |\Sigma_0|^2|h^+_1|^2  %+\lambda_6'  |\Phi|^2|\chi^+_2|^2
  +\lambda_{\Sigma h_2}  |\Sigma|^2|h^+_2|^2 +\lambda_{\Sigma k}  |\Sigma_0|^2|k^{++}|^2  +\lambda_{\Sigma \chi_0}  |\Sigma_0|^2|\chi_0|^2
 \nn\\
&
  + \lambda_{h_1} |h_1^{+}|^{4}
%+\lambda_6'  |\Phi|^2|\chi^+_2|^2
  {+} \lambda_{h_1 h_2}  |h_1^{+}|^2|h^+_2|^2 +\lambda_{h_1 k}  |h_1^{+}|^2|k^{++}|^2  +\lambda_{h_1 \chi_0}  |h_1^{+}|^2|\chi_0|^2
 \nn\\
&+
 \lambda_{h_2} |h_2^{+}|^{4} + \lambda_{h_2 k}  |h_2|^2|k^{++}|^2  +\lambda_{h_2 \chi_0}  |h_2^{+}|^2|\chi_0|^2
% \nn\\&
+ \lambda_{k} |k^{++}|^{4} 
+\lambda_{k \chi_0}  |k^{++}|^2|\chi_0|^2
,
\label{HP}
\end{align}
where the first term of $\mathcal{L}_{Y}$ generates the SM charged-lepton masses and $y_L$ ($y_R$) are three-by-three antisymmetric (symmetric) matrices, respectively.
We assume $\mu_{12}$, and $\mu_{22}$ to be positive real, but $\lambda_0$ to be negative real to identify $\chi_{0R}$ as the DM candidate {(see Eq.~(\ref{eq:mass_eigenvalues}))}.
%can be chosen to be real without any loss of generality by renormalizing the phases to scalar bosons. 
Here, we briefly mention the correspondence to the Zee-Babu model in the trilinear couplings among the charged scalars.
In the Zee-Babu model, only one singly-charged scalar is introduced and we regenerate the forms by taking the limits in our model, $h_{1}^{\pm} \to h^{\pm}$, $h_{2}^{\pm} \to h^{\pm}$, $\mu_{22} \to \mu$.

%%%
\subsection{Mass matrices of bosons}

The scalar fields can be parameterized as 
\begin{align}
%\begin{tiny}
&\Phi =\left[
\begin{array}{c}
w^+\\
\frac{v+\phi+i z}{\sqrt2}
\end{array}\right],\quad 
\
\chi_0=\frac{\chi_{0R}+i \chi_{0I}}{\sqrt{2}},
\ 
\Sigma_0=\frac{v'+\sigma}{\sqrt{2}}e^{iG/v'},
\label{component}
%\end{tiny}
\end{align}
where $v \simeq 246$ GeV is the VEV of the Higgs doublet field, and $w^\pm$
and $z$ are respectively {(would-be) NGB}
which are absorbed {as} the longitudinal {components} of $W$ and $Z$ bosons.
Inserting the tadpole conditions, $\partial\mathcal{V}/\partial\phi|_{v}=0$ and $\partial\mathcal{V}/\partial\sigma|_{v'}=0$,
{the} {resultant} mass matrix of the CP even bosons $(\phi,\sigma)$ 
 is given by
\begin{equation}
m^{2} (\phi,\sigma) = \left[%
\begin{array}{cc}
  2\lambda_\Phi v^2 & \lambda_{\Phi \Sigma}vv' \\
  \lambda_{\Phi \Sigma}vv' & 2\lambda_{\Sigma}v'^2 \\
\end{array}%
\right] = \left[\begin{array}{cc} \cos\alpha & \sin\alpha \\ -\sin\alpha & \cos\alpha \end{array}\right]
\left[\begin{array}{cc} m^2_{h} & 0 \\ 0 & m^2_{H}  \end{array}\right]
\left[\begin{array}{cc} \cos\alpha & -\sin\alpha \\ \sin\alpha &
      \cos\alpha \end{array}\right], 
\end{equation}
where $h$ is the SM-like Higgs and $H$ is an additional CP-even Higgs mass
eigenstate. The mixing angle $\alpha$ is {determined as} 
%$\sin 2\alpha=\frac{2\lambda_{13} v v'}{m^2_h-m_H^2}$.
\be
%\tan 2\alpha=\frac{\lambda_{13} v v'}{\lambda_{17} v'^2-\lambda_1 v^2}.
\sin 2\alpha=\frac{2\lambda_{\Phi \Sigma} v v'}{{m^2_H-m_h^2}}.
\label{eq:CP-even_mixing}
\ee
The Higgs bosons $\phi$ and $\sigma$ are rewritten in terms of the mass eigenstates $h$ and $H$ as
\begin{eqnarray}
\phi = h\cos\alpha + H\sin\alpha, %\nn\\
\quad
\sigma =- h\sin\alpha + H\cos\alpha.
\label{eq:mass_weak}
\end{eqnarray}
{An} NGB appears due to the spontaneous symmetry breaking of
the global $U(1)$ symmetry. 
%%%
The mass eigenvalues for the neutral bosons $\chi_{0R}$, $\chi_{0I}$,  the singly-charged bosons $h_1^\pm$, $h_2^\pm$ and the {doubly-charged} boson $k^{\pm\pm}$ are respectively given as
\begin{align}
&m^{2}_{\chi_{0R}} = m_{\chi_0}^{2}  + \frac{ \lambda_{\Phi\chi_0} v^{2}+({2}\lambda_{0}+\lambda_{\Sigma\chi_0}) v'^{2}}{2}, \quad 
%\\ &
  m^{2}_{\chi_{0I}} = m_{\chi_0}^{2}  + \frac{ \lambda_{\Phi\chi_0} v^{2}+(-{2}\lambda_{0}+\lambda_{\Sigma\chi_0}) v'^{2}}{2},  \notag \\ 
 %%%
&m^{2}_{h^{\pm}_1} = m_{h_1}^{2}  + \frac12 (\lambda_{\Phi h_1} v^{2}+\lambda_{\Sigma h_1} v'^{2}), \quad 
 %%%
m^{2}_{h^{\pm}_2} = m_{h_2}^{2}  + \frac12 (\lambda_{\Phi h_2} v^{2}+\lambda_{\Sigma h_2} v'^{2}), \notag \\ 
%%%
&m^{2}_{k^{\pm\pm}} = m_{k}^{2}  + \frac12 (\lambda_{\Phi k} v^{2}+\lambda_{\Sigma k} v'^{2}),
	\label{eq:mass_eigenvalues}
\end{align}
where these particles are not mixed due to the invariance of the system and thus they themselves are mass eigenstates, respectively.

\subsection{Vacuum stability of electrically charged bosons}

The vacuum stability has to be especially assured by the Higgs potential for {electrically-charged} bosons ($h_1^{\pm}, h_2^{\pm}, k^{\pm\pm}$). 
However, our model has some loop contributions to the leading order of these quartic couplings.
%, one has to take into considerration such a constraint.
Here, we {examine this issue} at the one-loop level. Let us define these quartic couplings as follows:
\begin{align}
&0\le\lambda_{h_1}=\lambda^{(0)}_{h_1}+\lambda^{(1)}_{h_1}, \notag \\
&0\le\lambda_{h_2}=\lambda^{(0)}_{h_2}+\lambda^{(1)}_{h_2}, \notag \\
&0\le\lambda_{k}=\lambda^{(0)}_{k}+\lambda^{(1)}_{k}, \label{cdt-lam1}
\end{align}
where the upper indices denote the number of the order, and the one-loop contributions can be given as
\begin{align}
&\lambda^{(1)}_{h_1}=
-\frac12 |\mu_{12}|^4 {\sum_{i=R,I} F_0(m_{h^{\pm}_2},m_{\chi_0(i)})},
	\label{quartic_lambda_1} \\
%%%
&\lambda^{(1)}_{h_2}=
-8 |\mu_{22}|^4 F_0(m_{h^{\pm}_2},m_{k^{\pm\pm}})-\frac12 |\mu_{12}|^4 {\sum_{i=R,I} F_0(m_{h^{\pm}_1},m_{\chi_0(i)})},
	\label{quartic_lambda_2} \\
%%%
&\lambda^{(1)}_{k}=
-4 |\mu_{22}|^4 F_0(m_{h^{\pm}_2},m_{h^{\pm}_2}),
	\label{quartic_lambda_3}
\end{align}
with
\begin{align}
&F_0(m_1,m_2)=\frac{1}{(4\pi)^2}\int_0^1 dx dy\delta(x+y-1)\frac{x y}{(x m_1^2+y m_2^2)^2},\label{cdt-vs}
\end{align}
where each of $m_1$ and $m_2$ of $F_0$ {represent} {a mass of} propagating fields in the loops.
We will include {these constraints} in the numerical analysis later.
%%%%
To avoid the global minimum {with electromagnetic charge-breaking} $\mathcal{V}(r\neq0)>0$, the following  condition should be at least satisfied:
\begin{align}
{2} |\mu_{12}+\mu_{22}| <  \sqrt{\Lambda}\left[m^2_\Phi + m^2_{h_1} + m^2_{h_2}+ m^2_{k}+ m^2_{\Sigma} + m^2_{\chi_0}   \right]^{1/2},%M,
\quad \Lambda\equiv\sum_{i={\rm all\ quartic\ couplings}}\lambda_i, 
%\quad M\equiv \left[\sum_{i={\rm all\ mass\ terms}}m_i^2\right]^{1/2},
\end{align}
where $r\equiv |\Phi|=|h^+_1|=|h^+_2|=|k^{++}|=|\Sigma_0|=|\chi_0|$.
If all these quartic couplings are of the order {as} $\lambda_i \approx {\mathcal{O}(\pi)}$~\footnote{{$\lambda_0$ is excluded, 
because $\lambda_0$ is negative and the maximum value is 0.}}, 
the following condition can be given by
%%%%
\begin{align}
|\mu_{12}+\mu_{22}| \  \lesssim \, {4.36} \sqrt{\pi} \left[m^2_{h_1} + m^2_{h_2}+ m^2_{k}+ m^2_{\Sigma} + m^2_{\chi_0}   \right]^{1/2},
	\label{eq:avoiding_chargedmim}
\end{align}
%%%%
where $m^2_\Phi$ and $\lambda_\Phi$ are neglected. 
{Note that the vacuum stability conditions take the following forms in the Zee-Babu model,
\begin{align}
&\lambda^{(1)}_{h_2} \to \lambda^{(1)}_{h}=
-8 |\mu|^4 F_0(m_{h^{\pm}},m_{k^{\pm\pm}}), \\
%%%
&\lambda^{(1)}_{k}=
-4 |\mu|^4 F_0(m_{h^{\pm}},m_{h^{\pm}}),
\end{align}
where no $\lambda^{(1)}_{h_1}$'s counterpart is there.
}

%\subsection{Neutrino mass matrix}
\subsection{Neutrino mass matrix}

%%%%%%%%%%%%%%%%%%%
\begin{figure}[t]
\begin{center}
\includegraphics[scale=0.7]{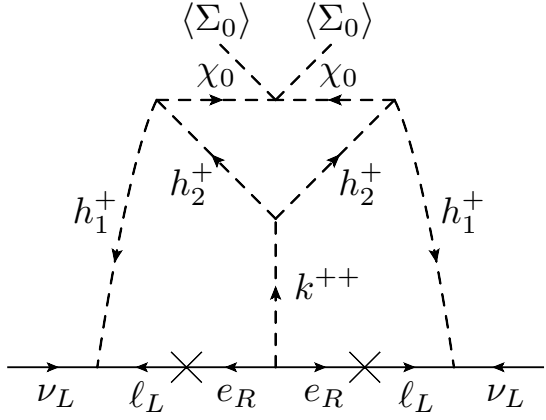}
   \caption{Radiative generation of neutrino masses.}
   \label{neutrino-diag}
\end{center}
\end{figure}
%%%%%%%%%%%%%%%%%%%
The Majorana neutrino mass matrix $m_\nu$ is derived at the three-loop level from the
diagrams depicted in Fig.~\ref{neutrino-diag}, which is described by an effective operator, $-\frac{1}{2} \overline{(\nu_{L_a})^c} (m_{\nu})_{ab} \nu_{L_b}$.
The concrete form of $(m_{\nu})_{ab}$ is given by
%\begin{widetext}
\begin{align}
&(m_{\nu})_{ab}
=
\frac{4\mu^2_{12}\mu_{22}}{(4\pi)^6 M^4%_{k^{\pm\pm}}
} %\sum_{\alpha=1}^{3} 
\sum_{i,j=1}^{3} %\sum_{\gamma=1}^{2}
% \sum_{\delta=1}^{3}
\left[
 %\frac{
 (y_L)_{ai}m_{\ell_i}
 (y^\dag_{R})_{ij}
 m_{\ell_j}
  (y^T_L)_{jb}
 \right] {\times}
	\label{eq:loop_fullform}
% \sin^2\theta \cos^2\theta \sin^22\theta
%\nn\\
%%%&\times 
\\
&\left[
F_1\left(
\frac{m_{h^+_1}^2}{M^2},%{m_{k^{\pm\pm}}^2},
\frac{m_{h^+_2}^2}{M^2},%{m_{k^{\pm\pm}}^2},
\frac{m_{\ell_i}^2}{M^2},%{m_{k^{\pm\pm}}^2},
\frac{m_{\ell_j}^2}{M^2},%{m_{k^{\pm\pm}}^2},
\frac{m_{\chi_{0R}}^2}{M^2},%{m_{k^{\pm\pm}}^2},
\frac{m_{k^{\pm\pm}}^2}{M^2}%{m_{k^{\pm\pm}}^2},
\right)
-
F_1\left(
\frac{m_{h^+_1}^2}{M^2},%{m_{k^{\pm\pm}}^2},
\frac{m_{h^+_2}^2}{M^2},%{m_{k^{\pm\pm}}^2},
\frac{m_{\ell_i}^2}{M^2},%{m_{k^{\pm\pm}}^2},
\frac{m_{\ell_j}^2}{M^2},%{m_{k^{\pm\pm}}^2},
\frac{m_{\chi_{0I}}^2}{M^2},%{m_{k^{\pm\pm}}^2},
\frac{m_{k^{\pm\pm}}^2}{M^2}%{m_{k^{\pm\pm}}^2},
\right)
\right],\nn
\end{align}%\end{widetext}
where $M={\rm max}[m_{k^{\pm\pm}},m_{h^\pm_1},m_{h^\pm_2},m_{\ell_{i/j}},m_{\chi_{0R}},m_{\chi_{0I}}]$ and the loop function $F_1$ is computed as
%\begin{widetext}
\begin{align} 
&F_1\left(X_1,X_2,X_3,X_4,X_5,X_6\right) 
=
\int d^3x\frac{\delta(x+y+z-1)}{y(y-1)+z(z-1)+2yz}\nn \\
%%%
&\times
\int d^4x'\frac{\delta(\alpha+\beta+\gamma+\delta-1)}{((\alpha Y+\delta)^2-\delta-\alpha Y^2-\alpha X)^2}%
%%%
\nn\\&
\times
\int d^3x''\frac{\rho\delta(\rho+\sigma+\omega-1)}
{\left[ \rho A\left(X_1, X_2, X_3, X_5, X_6\right) -\sigma X_4-\omega X_1\right]^2},
	\label{eq:threeloop_function}
\end{align}
%\end{widetext}
with 
\begin{align}
A\left(X_1,X_2,X_3,X_5,X_6 \right) &=
-\frac{\alpha((x+y) X_2 + z X_5)}{((\alpha Y+\delta)^2-\delta-\alpha Y^2-\alpha X)(y(y-1)+z(z-1)+2yz)}
\nn\\&
+
\frac{\beta X_1 +\gamma X_3+ \delta X_6}{((\alpha Y+\delta)^2-\delta-\alpha Y^2-\alpha X)},\\
%%%
X&=-\left(\frac{y}{y+z}\right)^2+\frac{y(y-1)}{y(y-1)+z(z-1)+2yz},\quad
Y=\frac{y}{y+z},
\end{align}
where we define $\int d^3x\delta(x+y+z-1)\equiv \int_{0}^{1}dx\int_{0}^{1-x}dy$, $\int {d^4x'} \delta(\alpha+\beta+\gamma+\delta-1)\equiv  \int_{0}^{1}d\alpha  \int_{0}^{1-\alpha}d\beta  \int_{0}^{1-\alpha-\beta} d\gamma$, and  $\int d^3x''\delta(\rho+\sigma+\omega-1) \equiv  \int_{0}^{1} d\rho
 \int_{0}^{1-\rho}d\sigma$~\footnote{{We assume $m_{\ell_{i/j}}\approx 0$} in our numerical analysis, since these masses are much smaller than the other masses {inside the loops}.}. 
The neutrino mass {eigenstates} and their {mixings} can be straightforwardly given by applying {them} to the Zee-Babu analogy~\cite{Herrero-Garcia:2014hfa}, since the structure of the fermion line is the same as the the Zee-Babu model~\cite{zee-babu}, that is, a rank two model of the neutrino mass matrix due to the {antisymmetricity} of $y_L$.
Let us define the neutrino mass matrix as   
\begin{align}
&(m_{\nu})_{ab}=(U_{{\text{PMNS}}} m^{diag}_{\nu} U^T_{{\text{PMNS}}})_{ab}\equiv \zeta (y_L)_{ai}\omega_{ij} (y^T_L)_{jb},
	\label{eq:neutrino_massmatrix} \\
&\zeta=\frac{4\mu^2_{12}\mu_{22}}{(4\pi)^6 M^4} 
\left[
F_1\left(X_{iR}\right) - F_1\left(X_{iI}\right)\right],
	\label{eq:neutrino_loopfactor}\\
& \omega_{ij}=m_{\ell_i}(y^\dag_R)_{ij}m_{\ell_j},
	\label{eq:definition_omega}
\end{align}
where $i$ runs over $1$ to {$3$},
$m^{diag}_{\nu}\equiv (m_1,m_2,m_3)$ are the neutrino mass eigenvalues, and $U_{{\text{PMNS}}}$ ({Pontecorvo-}Maki-{Nakagawa}-Sakata matrix~\cite{Maki:1962mu,Pontecorvo:1967fh}) is the mixing matrix to diagonalize the neutrino mass matrix, {which is {parametrized} as}~\cite{Herrero-Garcia:2014hfa}
\begin{align}
U_{{\text{PMNS}}}=
\left[\begin{array}{ccc} {c_{13}}c_{12} &c_{13}s_{12} & s_{13} e^{-i\delta}\\
 -c_{23}s_{12}-s_{23}s_{13}c_{12}e^{i\delta} & c_{23}c_{12}-s_{23}s_{13}s_{12}e^{i\delta} & s_{23}c_{13}\\
  s_{23}s_{12}-c_{23}s_{13}c_{12}e^{i\delta} & -s_{23}c_{12}-c_{23}s_{13}s_{12}e^{i\delta} & c_{23}c_{13}\\
  \end{array}
\right]
%%%
\left[\begin{array}{ccc} 1 & 0 & 0   \\
0 &e^{i\phi/2} & 0\\
 0 & 0 & 1\\
  \end{array}
\right],
\end{align}
where $c_{ij}\equiv \cos\theta_{ij}$ and $s_{ij}\equiv \sin\theta_{ij}$ with $(i,j)=(1-3)$.
Depending on the ordering of the neutrino masses, whether normal ($m_1\,(=0) < m_2 < m_3$) or inverted ($m_3\,(=0) < m_1 < m_2$) in our case, one can derive some simple formulae~\footnote{More details {are} given in Ref.~\cite{Herrero-Garcia:2014hfa} for both cases.}.
%%%%%%%%%%%%%
{When we consider} the normal ordering,
{the following relations should hold for realizing the observed neutrino profiles,}
\begin{align}
&y_{L_{13}} = (s_{12} c_{23}/(c_{12} c_{13}) + s_{13} s_{23} e^{-i\delta}/c_{13}) y_{L_{23}}, \notag \\
&y_{L_{12}} = (s_{12} s_{23}/(c_{12} c_{13}) - s_{13} c_{23} e^{-i\delta}/c_{13}) y_{L_{23}}, \notag \\
%%%
&\zeta y_{L_{23}}^2 \omega_{33}\approx m_3 c^2_{13}s^2_{23}+m_2 e^{i\phi}(c_{12}c_{23}-e^{i\delta}s_{12}s_{13}s_{23})^2, \notag \\
&\zeta y_{L_{23}}^2 \omega_{23}\approx -m_3 c^2_{13}c_{23}s_{23}
+m_2 e^{i\phi}(c_{12}s_{23}+e^{i\delta}c_{23}s_{12}s_{13})(c_{12}c_{23}-e^{i\delta}s_{12}s_{13}s_{23}), \notag \\
&\zeta y_{L_{23}}^2 \omega_{22}\approx m_3 c^2_{13}c^2_{23}+m_2 e^{i\phi}(c_{12}s_{23} {+} e^{i\delta}c_{23}s_{12}s_{13})^2,
	\label{eq:normal_ordering}
\end{align}
where we use $m_e \,{\ll}\, m_\mu, m_\tau$.
In the case of the inverted neutrino mass hierarchy, the conditions are deformed as
%%%%
\begin{align}
&y_{L_{13}} = - (c_{13} s_{23} e^{-i\delta}/s_{13}) y_{L_{23}}, \notag \\
&y_{L_{12}} = + (c_{13} c_{23} e^{-i\delta}/s_{13}) y_{L_{23}}, \notag \\
%%%
&\zeta y_{L_{23}}^2 \omega_{33} \approx m_1 (c_{23}s_{12} + e^{i\delta}c_{12}s_{13}s_{23})^2 + m_2 e^{i\phi}(c_{12}c_{23} - e^{i\delta}s_{12}s_{13}s_{23})^2, \notag \\
&\zeta y_{L_{23}}^2 \omega_{23} \approx m_1 (s_{12}s_{23} - e^{i\delta}c_{12}c_{23}s_{13})(c_{23}s_{12} + e^{i\delta}c_{12}s_{13}s_{23}) \notag \\
&\qquad \qquad \quad + m_2 e^{i\phi}(c_{12}s_{23}+e^{i\delta}c_{23}s_{12}s_{13})(c_{12}c_{23}-e^{i\delta}s_{12}s_{13}s_{23}), \notag \\
&\zeta y_{L_{23}}^2 \omega_{22} \approx m_1 (s_{12}s_{23} - e^{-\delta}c_{12}c_{23}s_{13})^2 + m_2 e^{i\phi}(c_{12}s_{23} {+} e^{i\delta}c_{23}s_{12}s_{13})^2.
	\label{eq:inverse_ordering}
\end{align}
Here, we mention that these conditions take the same forms in the Zee-Babu model up to the contexts of the loop factor $\zeta$ in Eq.~(\ref{eq:neutrino_massmatrix}).

\subsection{Lepton flavor violations and the universality of charged currents}

In our model, there exist several lepton flavor violating processes and the universality violation of charged currents even at tree level order.
They put some constraints on the parameter spaces.
Since all the processes are exactly {the same with} the ones of the Zee-Babu model~\cite{Herrero-Garcia:2014hfa} {after the replacement of $h^{\pm}$ as $h_1^{\pm}$}, we just list up such kind of bounds below.
\begin{align}
%%%%  tree level contributions 
& |{y}_{R_{12}} {y}^*_{R_{11}}|< 2.3\times 10^{-5}\ \left(\frac{{m_{k^{\pm\pm}}}}{{\rm TeV}}\right)^2,\
 |{y}_{R_{13}} {y}^*_{R_{11}}|< 0.009\ \left(\frac{{m_{k^{\pm\pm}}}}{{\rm TeV}}\right)^2, \notag \\
& |{y}_{R_{13}} {y}^*_{R_{12}}|< 0.005\ \left(\frac{{m_{k^{\pm\pm}}}}{{\rm TeV}}\right)^2,\
 |{y}_{R_{13}} {y}^*_{R_{22}}|< 0.007\ \left(\frac{{m_{k^{\pm\pm}}}}{{\rm TeV}}\right)^2, \notag\\
& |{y}_{R_{23}} {y}^*_{R_{11}}|< 0.007\ \left(\frac{{m_{k^{\pm\pm}}}}{{\rm TeV}}\right)^2,\
 |{y}_{R_{23}} {y}^*_{R_{12}}|< 0.007\ \left(\frac{{m_{k^{\pm\pm}}}}{{\rm TeV}}\right)^2, \notag \\
& |{y}_{R_{23}} {y}^*_{R_{22}}|< 0.008\ \left(\frac{{m_{k^{\pm\pm}}}}{{\rm TeV}}\right)^2,\
 |{y}_{R_{11}} {y}^*_{R_{22}}|< 0.2\ \left(\frac{{m_{k^{\pm\pm}}}}{{\rm TeV}}\right)^2, \notag \\
	\label{eq:constraint_1}
\end{align}
\vspace{-8mm}
\begin{align}
%%%%  universality of the charged currents 
& |{y}_{L_{12}}|^2 <  0.007\ \left(\frac{{m_{h_{1}^\pm}}}{{\rm TeV}}\right)^2,
 ||{y}_{L_{{23}}}|^2-|{y}_{L_{13}}|^2| <  0.024 \ \left(\frac{{m_{h_{1}^\pm}}}{{\rm TeV}}\right)^2, \notag \\
& ||{y}_{L_{13}}|^2-|{y}_{L_{12}}|^2| <  0.035 \ \left(\frac{{m_{h_{1}^\pm}}}{{\rm TeV}}\right)^2,\
 ||{y}_{L_{23}}|^2-|{y}_{L_{12}}|^2| <  0.04 \ \left(\frac{{m_{h_{1}^\pm}}}{{\rm TeV}}\right)^2,
	\label{eq:constraint_2}
\end{align}
\vspace{-8mm}
\begin{align}
%%%%  one-loop  level contributions 
& {r^2} |{y}^*_{L_{13}} {y}_{L_{23}}|^2 + 16 |{y}^*_{R_{11}} {y}_{R_{12}}  +{y}^*_{R_{12}} {y}_{R_{22}}  + {y}^*_{R_{13}} {y}_{R_{23}}|^2 < 1.6\times 10^{-6}
\ \left(\frac{{m_{k^{\pm\pm}}}}{{\rm TeV}}\right)^{{4}},\notag \\
& {r^2} |{y}^*_{L_{12}} {y}_{L_{23}}|^2 + 16 |{y}^*_{R_{11}} {y}_{R_{13}}  +{y}^*_{R_{12}} {y}_{R_{23}}  + {y}^*_{R_{13}} {y}_{R_{33}}|^2 < 0.52\ \left(\frac{{m_{k^{\pm\pm}}}}{{\rm TeV}}\right)^{{4}},\notag \\
& {r^2} |{y}^*_{L_{12}} {y}_{L_{13}}|^2 + 16 |{y}^*_{R_{12}} {y}_{R_{13}}  +{y}^*_{R_{22}} {y}_{R_{23}}  + {y}^*_{R_{23}} {y}_{R_{33}}|^2 < 0.7\ \left(\frac{{m_{k^{\pm\pm}}}}{{\rm TeV}}\right)^{{4}},
	\label{eq:constraint_3}
\end{align}
where $r\equiv ({m_{k^{\pm\pm}}}/{m_{h_{1}^\pm}})^2$ {and the constraints in Eqs.~(\ref{eq:constraint_1}), (\ref{eq:constraint_2}) and (\ref{eq:constraint_3}) originate from (tree-level) lepton flavor violating decays of charged leptons, charged lepton gauge universalities and (loop-level) lepton flavor violating interactions associating with photon, respectively.}

\subsection{Numerical {analysis}}

Here, we have numerical analysis on our model and also the Zee-Babu model considering all the above constraints, namely, the vacuum stability of the three charged scalars in Eq.~(\ref{cdt-lam1}), avoiding charge-breaking minimum in Eq.~(\ref{eq:avoiding_chargedmim}), the observed neutrino masses and the mixings in Eq.~(\ref{eq:normal_ordering}) or (\ref{eq:inverse_ordering}), the lepton flavor violating processes and gauge universality in the charged leptons in Eqs.~(\ref{eq:constraint_1}), (\ref{eq:constraint_2}) and (\ref{eq:constraint_3}).
In addition, we add the following two conditions: (i) the mass of the DM candidate, {$\chi_{0R}$}, takes the smallest value among the particles with negative $\mathbb{Z}_2$ parity; (ii) all the quartic couplings at the one-loop level in Eq.~(\ref{cdt-lam1}), $\lambda_{h_1}$, $\lambda_{h_2}$ and $\lambda_{k}$, should be less than $\pi$ to ensure the perturbativity to a reasonable extent.

We fix and take the following parameter ranges:
\begin{align}
&\lambda^{(0)}_{h_1}\approx \lambda^{(0)}_{h_2}\approx \lambda^{(0)}_{k}\approx \pi,\  ({y}_{R_{11}})= ({y}_{R_{12}})=
 ({y}_{R_{13}})\approx 0,\  -\pi\le {y}_{L_{23}}\le \pi, \notag \\
& 0\le \delta\le \pi,\ 0\le \phi\le \pi,\ \mu_{12}\approx \mu_{22}\approx 10^5\ {\rm GeV}, \notag \\
%%%
&0\ {\rm GeV}\le m_{h_1^{\pm}} ,\  m_{h_2^{\pm}} ,\  m_{k^{\pm\pm}} ,\   m_{\chi_{0R/I}} \le 1.2\times10^5\ {\rm GeV} \notag \\
&\qquad \quad (\text{for our model in NH, Zee-Babu model in NH and IH}), \notag \\
%%%
&0\ {\rm GeV}\le m_{h_1^{\pm}} ,\  m_{h_2^{\pm}} ,\  m_{k^{\pm\pm}} ,\   m_{\chi_{0R/I}} \le 2.0\times10^5\ {\rm GeV} \notag \\
&\qquad \quad (\text{for our model in IH}),
%%%
\label{paraset}
%\label{para-set}
\end{align}
where NH and IH are short-hand notations of normal and inverted hierarchies, respectively.
In our three-loop situation, neutrino masses tend to be suppressed significantly because of the large three-loop suppression factor, where we remember that elements of $y_L$ and $y_R$ should not be so large to be consistent with the constraints in Eqs.~(\ref{eq:constraint_1}), (\ref{eq:constraint_2}) and (\ref{eq:constraint_3}).
Thereby, the option is assigning huge numbers in $\mu_{12}$ and $\mu_{22}$, as apparent from Eqs.~(\ref{eq:neutrino_massmatrix}) and (\ref{eq:neutrino_loopfactor}), which enhances realized values of neutrino masses.
Note that in the above choice, we safely avoid charge-breaking minimum when the masses of the scalars are compatible (or more) compared with $\mu_{12}$ and $\mu_{22}$. Then we focus on the other conditions in the scanning.
Here, we remember that the trilinear couplings among the charged scalars, $\mu_{12}$ and $\mu_{22}$, should be very large for generating enough amounts of neutrino masses.
From Eqs.~(\ref{quartic_lambda_1})--(\ref{quartic_lambda_3}), we notice that the largeness in $\mu_{12}$ and $\mu_{22}$ possibly endangers the vacuum stability due to negative quartic couplings at the one-loop level.
To maintain the stability of these couplings, three charged scalars should be suitably heavy.

\begin{figure}[tbc]
\begin{center}
\includegraphics[scale=0.55]{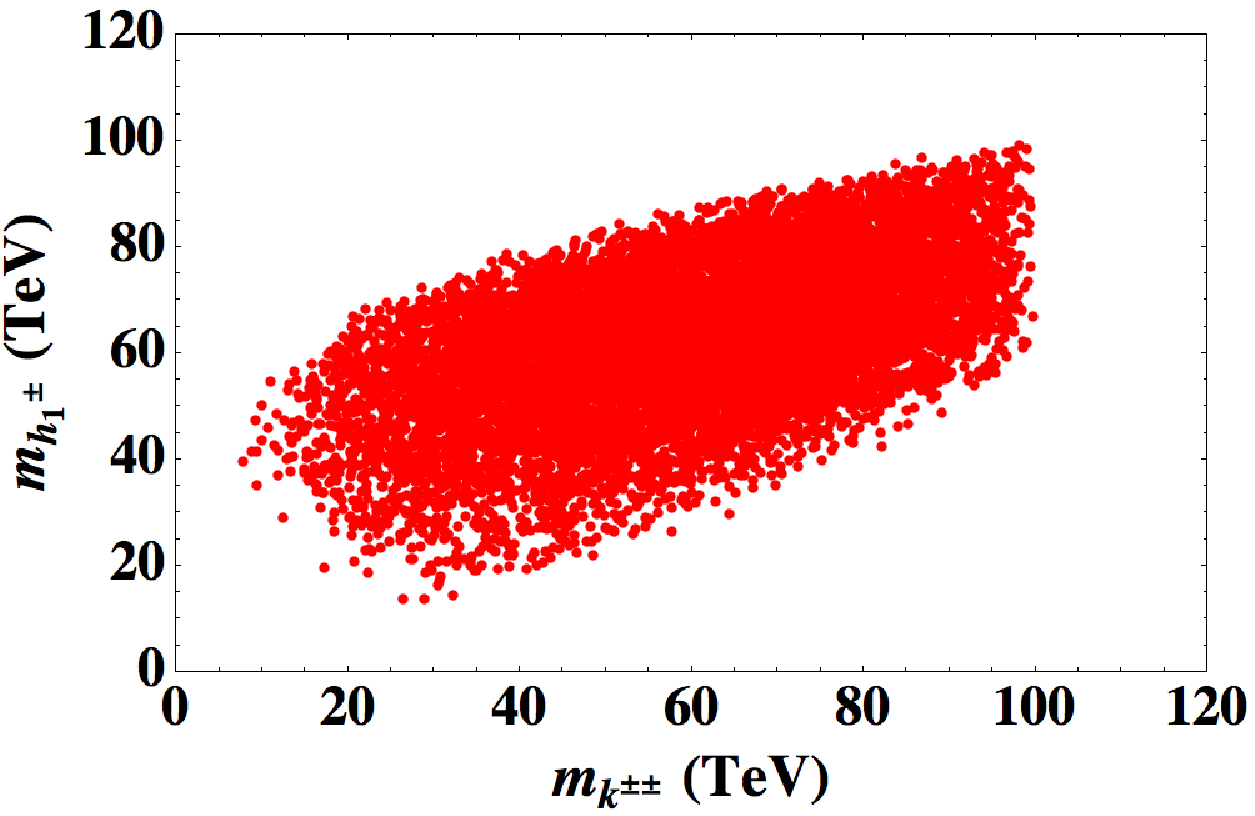}
\includegraphics[scale=0.55]{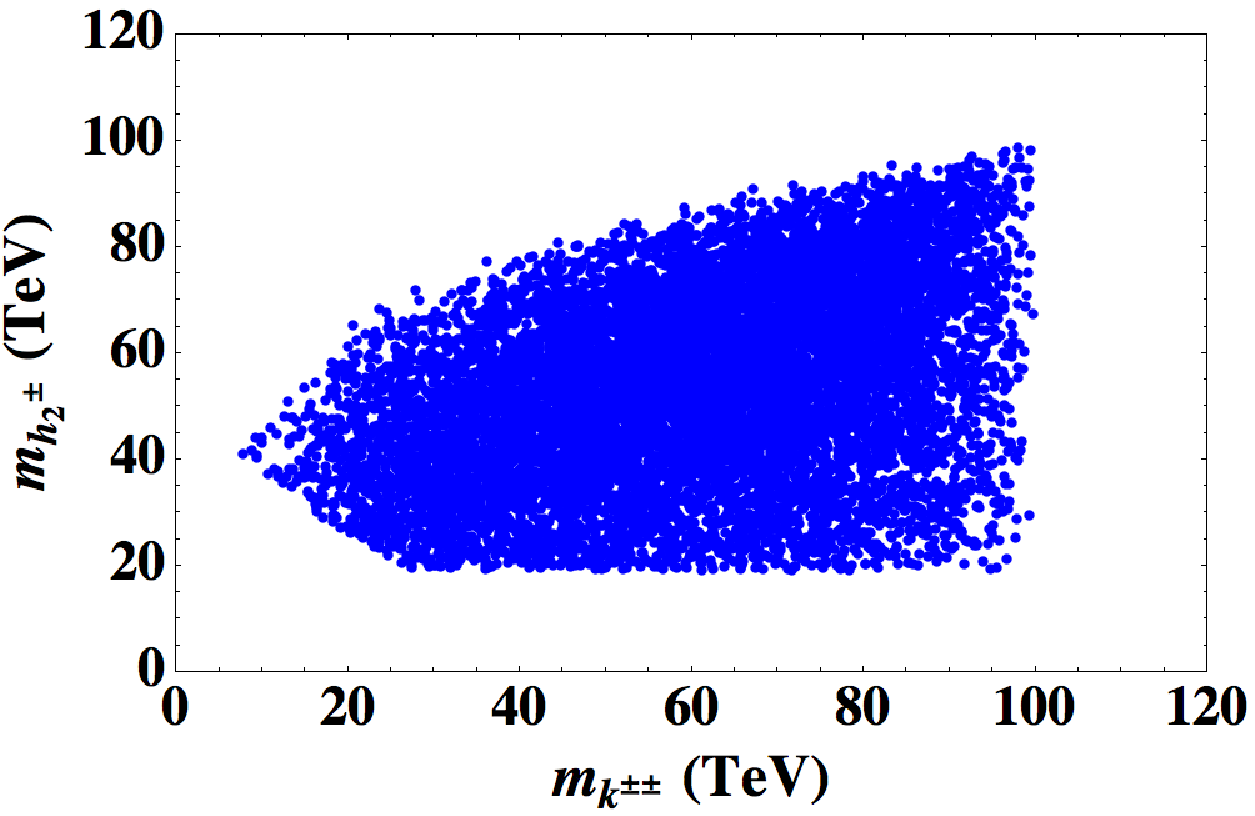}
\includegraphics[scale=0.55]{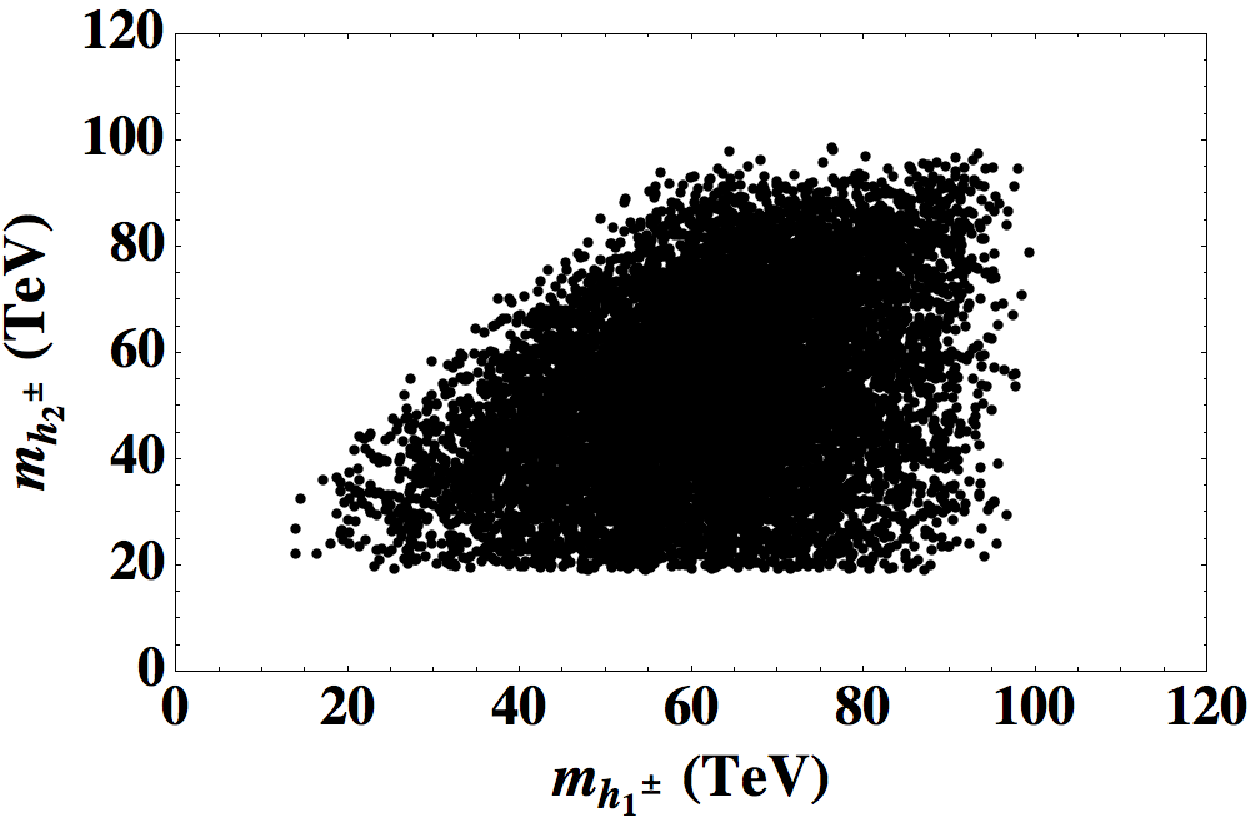}
%%%%%%%
\caption{The allowed mass {ranges} in the {NH} for the {singly-charged} bosons $(h_1^\pm, h_2^\pm)$ and the doubly-charged boson $k^{\pm\pm}$ to satisfy the vacuum stability of these charged bosons,  all the lepton flavor violating processes, the observed neutrino masses and the {mixings} under our parameter set in Eq.~(\ref{paraset}).
 These figures {tell} us {$10\ \text{TeV} \lesssim {m_{k^{\pm\pm}}} \lesssim 100\ \text{TeV}$, $10\ \text{TeV} \lesssim {m_{h_{1}^\pm}} \lesssim 100\ \text{TeV}$, and $20\ \text{TeV} \lesssim {m_{h_{2}^\pm}} \lesssim 100\ \text{TeV}$} are respectively allowed.
Note that we examine $10^6$ points in this scanning.
}
   \label{fig:mass_NH_cocktail}
\end{center}
\end{figure}

\begin{figure}[tbc]
\begin{center}
\includegraphics[scale=0.55]{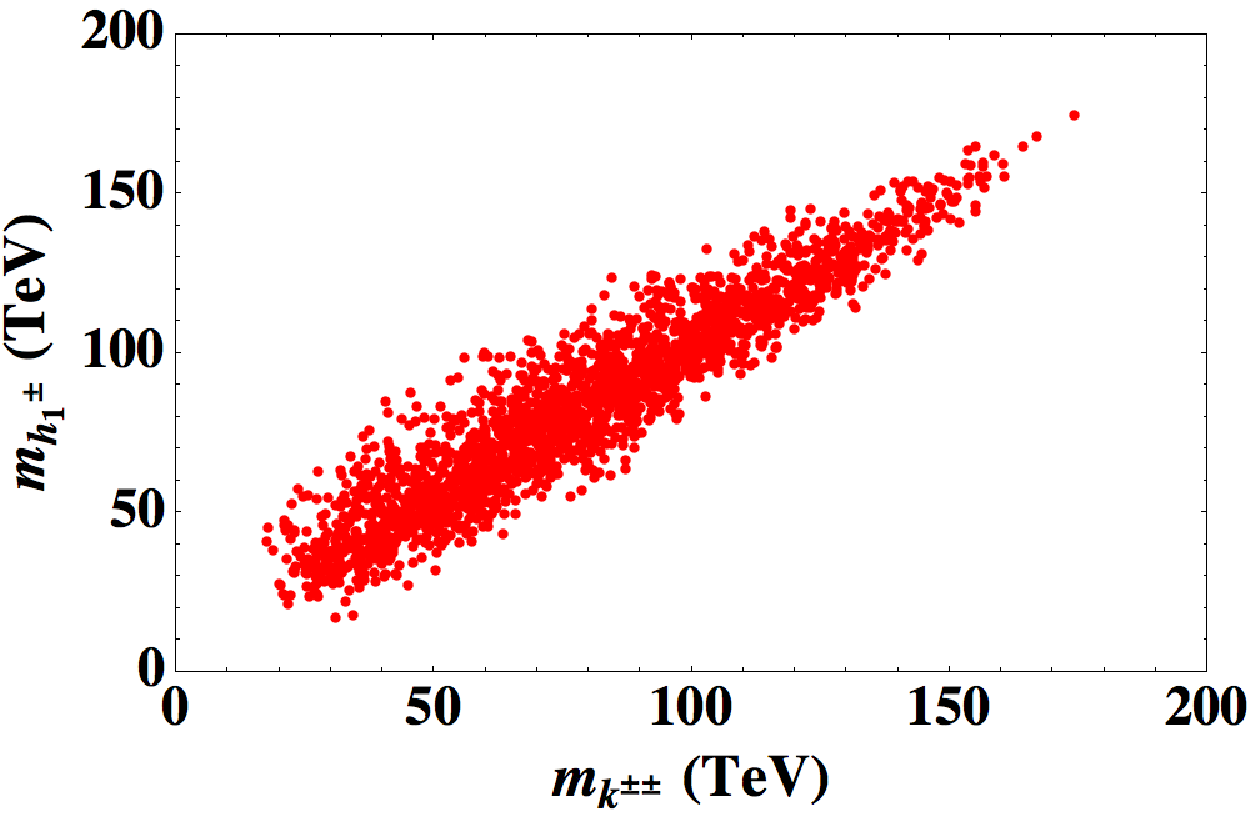}
\includegraphics[scale=0.55]{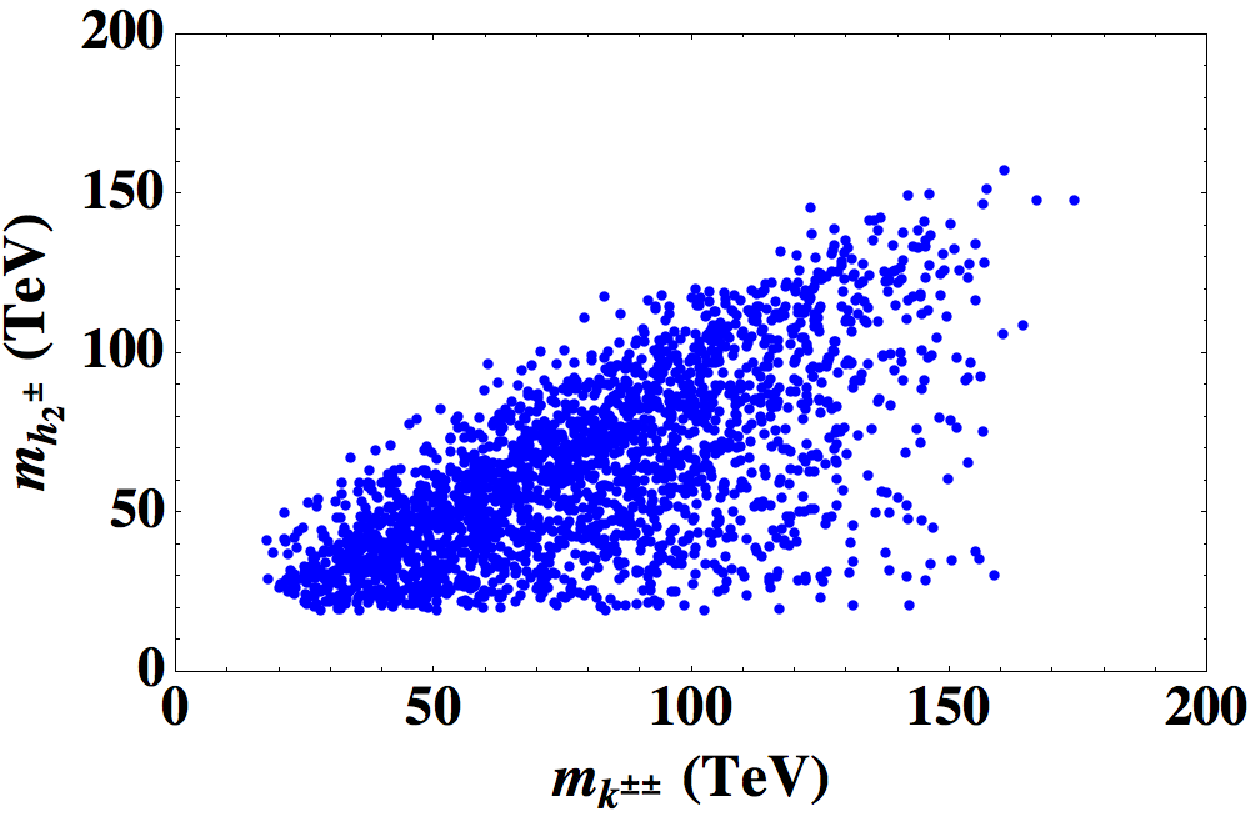}
\includegraphics[scale=0.55]{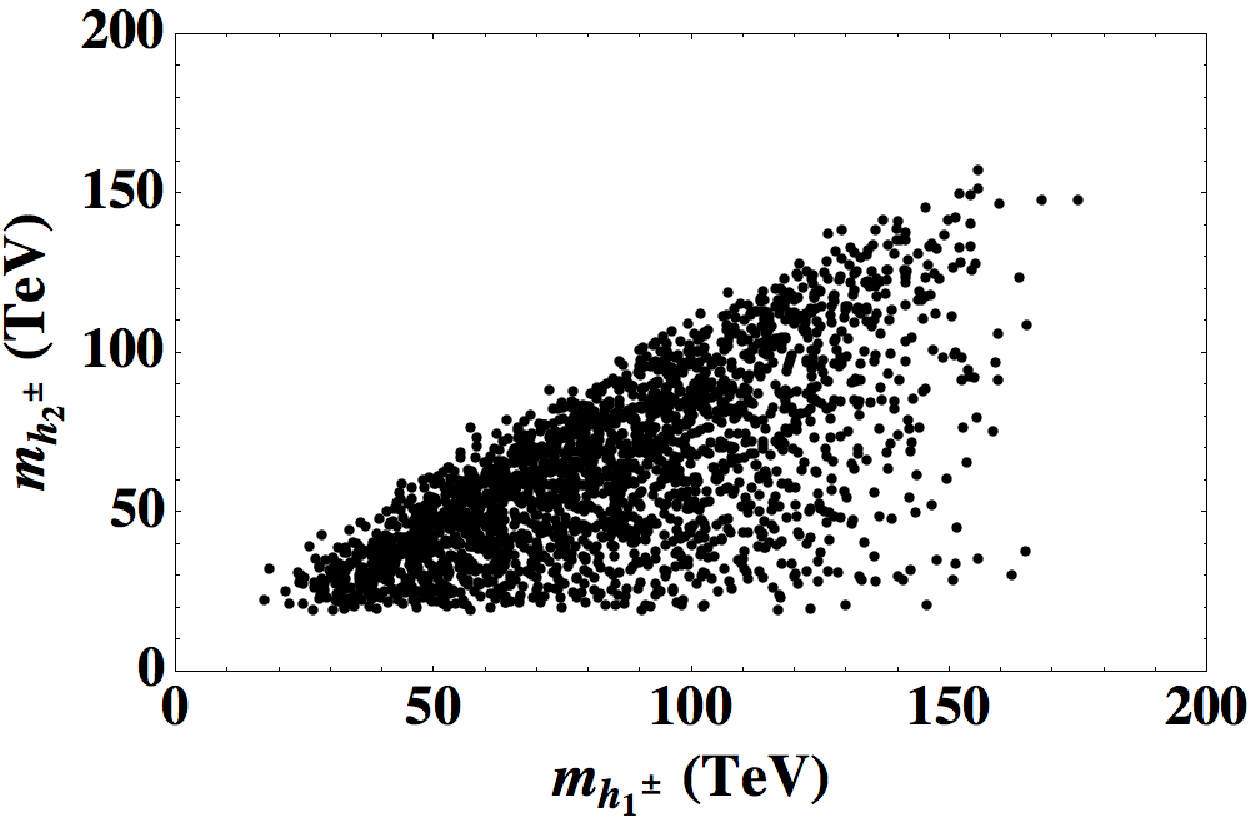}
%%%%%%%
\caption{The allowed mass {ranges} in the {IH} for the {singly-charged} bosons $(h_1^\pm, h_2^\pm)$ and the doubly-charged boson $k^{\pm\pm}$ to satisfy the vacuum stability of these charged bosons,  all the lepton flavor violating processes, the observed neutrino masses and the {mixings} under our parameter set in Eq.~(\ref{paraset}).
These figures tell us {$10\ \text{TeV} \lesssim {m_{k^{\pm\pm}}} \lesssim 170\ \text{TeV}$, $20\ \text{TeV} \lesssim {m_{h_{1}^\pm}} \lesssim 170\ \text{TeV}$, and $20\ \text{TeV} \lesssim {m_{h_{2}^\pm}} \lesssim 150\ \text{TeV}$} are respectively allowed. 
Note that we examine $10^8$ points in this scanning.
}
   \label{fig:mass_IH_cocktail}
\end{center}
\end{figure}

\begin{figure}[tbc]
\begin{center}
\includegraphics[scale=0.55]{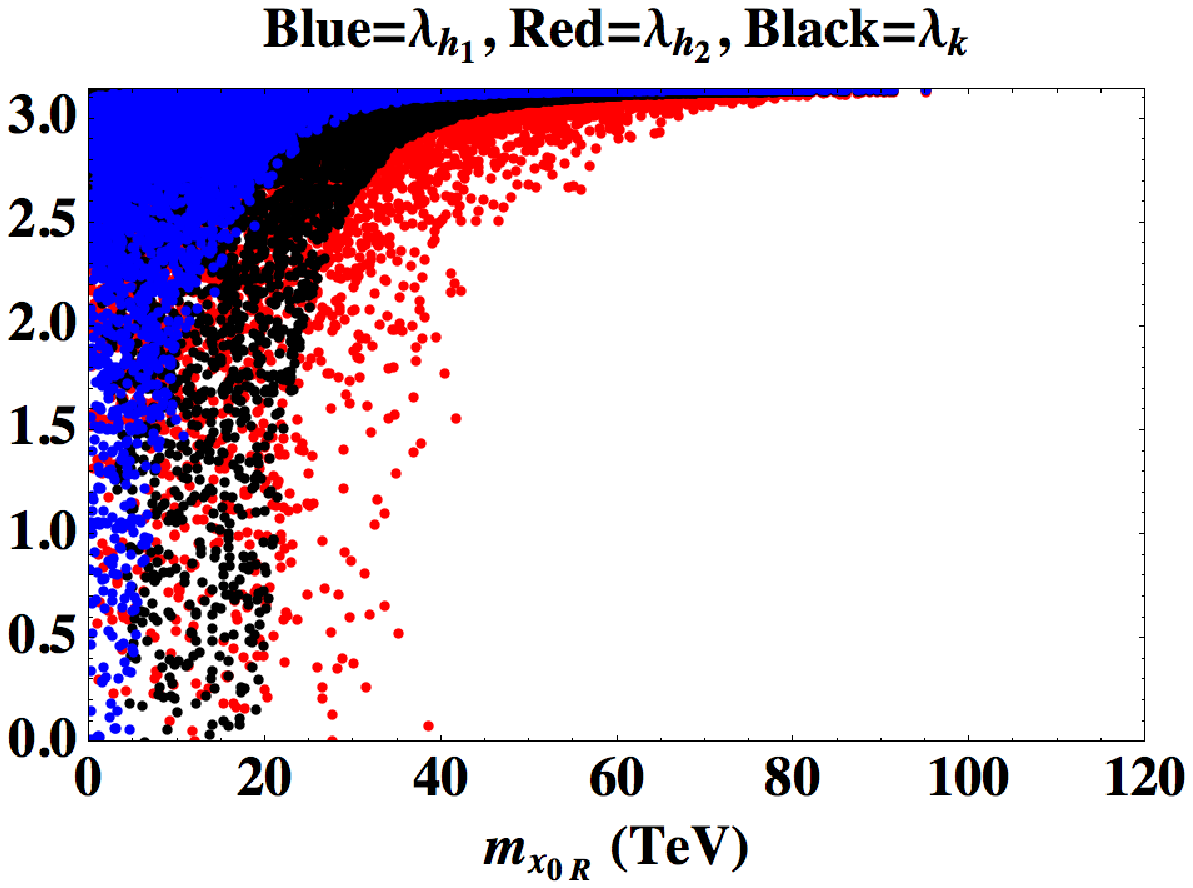}
\includegraphics[scale=0.55]{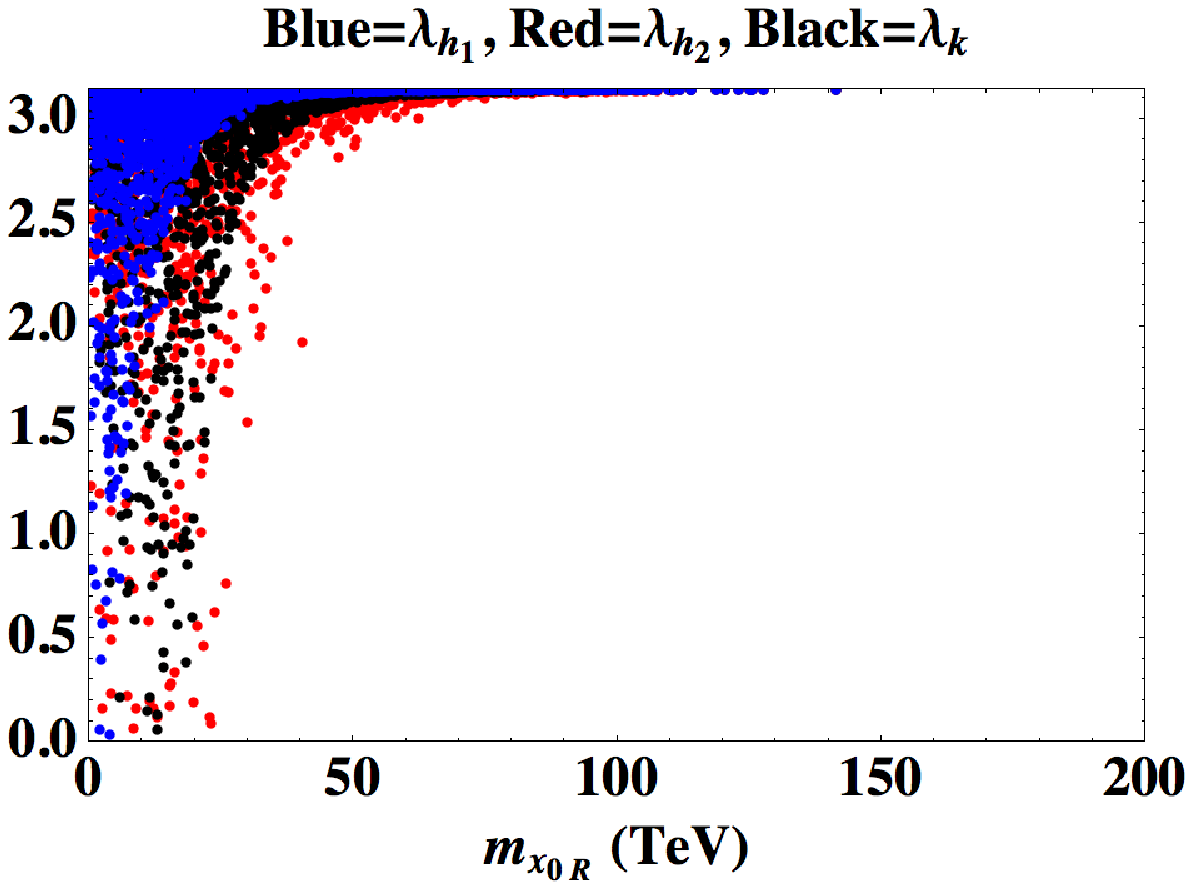}
%%%%
\caption{{The allowed range of quartic couplings for charged bosons ($\lambda_{h_1}$, $\lambda_{h_2}$, $\lambda_k$) and the DM mass, where the left figure is the {NH} case, while the right one is the {IH} case.}
Note that we examine $10^6$ ($10^8$) points in the {NH} ({IH}), respectively.
}
   \label{fig:dm}
\end{center}
\end{figure}

We can check that the three-loop function $F_1$ defined in Eq.~(\ref{eq:threeloop_function}) typically generates $\mathcal{O}(1)$ values in most of the part of the parameter space.
Therefore, we set the loop function part, $\left[F_1\left(X_{iR}\right) - F_1\left(X_{iI}\right)\right]$ in Eq.~(\ref{eq:neutrino_loopfactor}), as $0.625$ in scanning as a typical value.
{We search for suitable points within $10^6$, $10^8$ and $10^5$ candidates via the parameter landscape defined in Eq.~(\ref{paraset}) in the cases of our model in NH; our model in IH; Zee-Babu model in both of NH and IN, respectively.}
The results that satisfy all the data discussed above are found in Figures~\ref{fig:mass_NH_cocktail},~\ref{fig:mass_IH_cocktail}  and~\ref{fig:dm}.
%%%%%%%%%%

 Figure~\ref{fig:mass_NH_cocktail} shows the allowed mass ranges in the {NH case} for the {singly-charged} bosons $(h^{\pm}_1, h^{\pm}_2)$ and the doubly-charged boson {$k^{\pm\pm}$} to satisfy the vacuum stability of these charged bosons,  all the lepton flavor violating processes, universality of the charged currents, the observed neutrino masses and the {mixings} under our parameter set in Eq.~(\ref{paraset}). These figures tell us {$10\ \text{TeV} \lesssim {m_{k^{\pm\pm}}} \lesssim 100\ \text{TeV}$, $10\ \text{TeV} \lesssim {m_{h_{1}^\pm}} \lesssim 100\ \text{TeV}$, and $20\ \text{TeV} \lesssim {m_{h_{2}^\pm}} \lesssim 100\ \text{TeV}$} are respectively allowed.
 
%%%%%%%%%%

 Figure~\ref{fig:mass_IH_cocktail} shows the allowed mass ranges in the {IH case} for the {singly-charged} bosons $(h^{\pm}_1, h^{\pm}_2)$ and the doubly-charged boson {$k^{\pm\pm}$} to satisfy all the constraints discussed in the {NH} case.
Now, these figures tell us that {$10\ \text{TeV} \lesssim {m_{k^{\pm\pm}}} \lesssim 170\ \text{TeV}$, $20\ \text{TeV} \lesssim {m_{h_{1}^\pm}} \lesssim 170\ \text{TeV}$, and $20\ \text{TeV} \lesssim {m_{h_{2}^\pm}} \lesssim 150\ \text{TeV}$} are respectively allowed. Comparing to the {NH} case, one finds that heavier charged particles are allowed.
On the other hand, the allowed region in the parameter space is decreased, which is recognized via the densities of the points showing allowed configurations.
Note that the numbers of scanned points are different between the normal case $(10^6)$ and the inverted one $(10^8)$.
 %%%%%%%

%%%
Figure~\ref{fig:dm} shows the quartic couplings of $\lambda_{h_1,h_2,k}$ in terms of  the allowed mass range  of the DM candidate $m_{\chi_{0R}}$ for both ordering cases, in which one finds that the allowed region for the {IH} decreases drastically than the one for the {NH}. 
Since the mass of the DM should be the smallest among the particles with negative $\mathbb{Z}_2$ parity, its value is bounded from above via the allowed mass range of $m_{h_2^{\pm}}$ typically as $m_{h_2^{\pm}} \lesssim \mathcal{O}(10^2)\ \text{TeV}$.
%To maintain the stability of these couplings, the DM mass is expected not to be so large that is ${\cal O}$(100) TeV, unless the charged boson masses become much heavier or take the constraint of the vacuum stability more loosen in Eqs.~(\ref{cdt-lam1})--(\ref{cdt-vs}).
In this sense, our DM can naturally explain the observed relic density~\cite{Ade:2013zuv} and the direct detection searches~\cite{Akerib:2013tjd} which typically lies on the ${\cal O}$(100) GeV mass scale. The details of the DM properties can be found in Ref.~\cite{Baek:2013fsa}, just  replacing $m_5\to v'/\sqrt2$~\footnote{Since the lower bound of the DM mass is assumed to be larger than the tau lepton mass $\approx$ 1 GeV to simplify the neutrino mass formula in Eq.~(\ref{eq:loop_fullform}), the typical order could be larger than ${\cal O}$(10) GeV.}.

\begin{figure}[tbc]
\begin{center}
\includegraphics[scale=0.55]{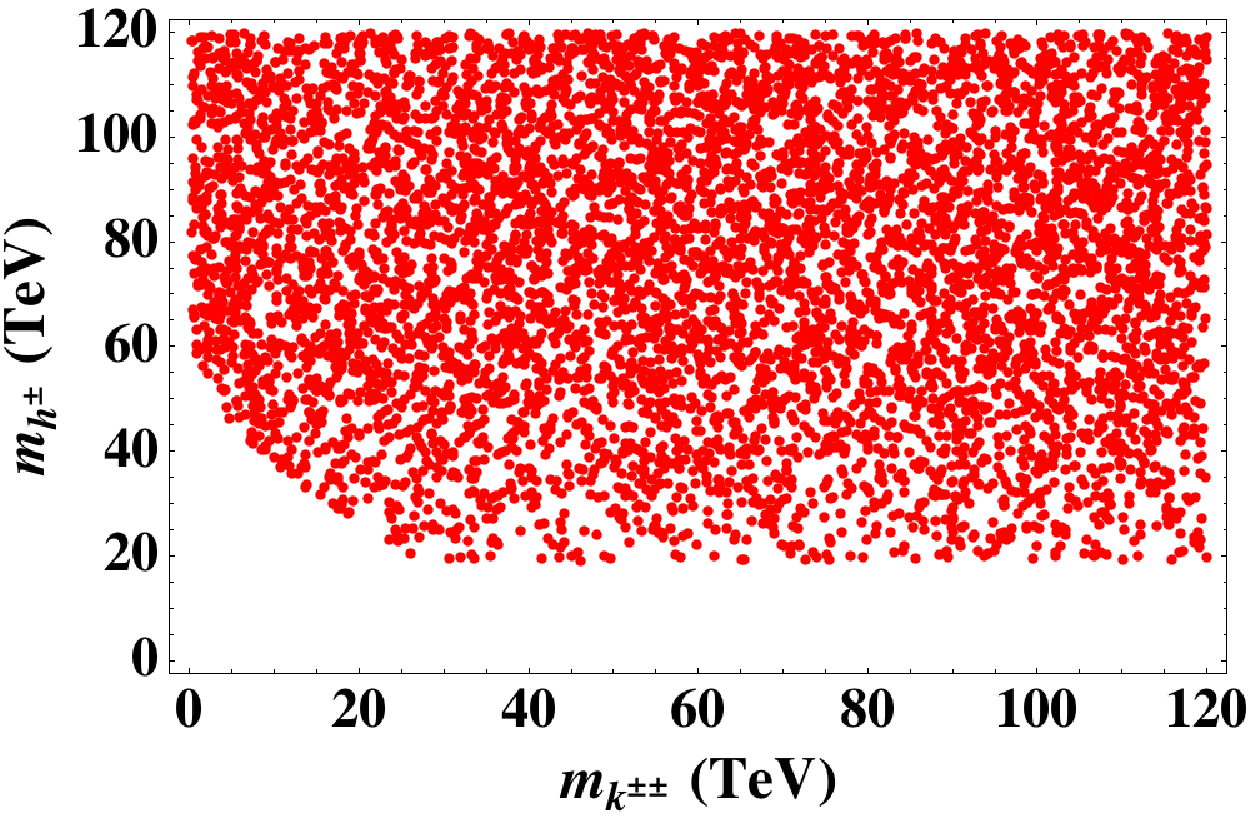}
\includegraphics[scale=0.55]{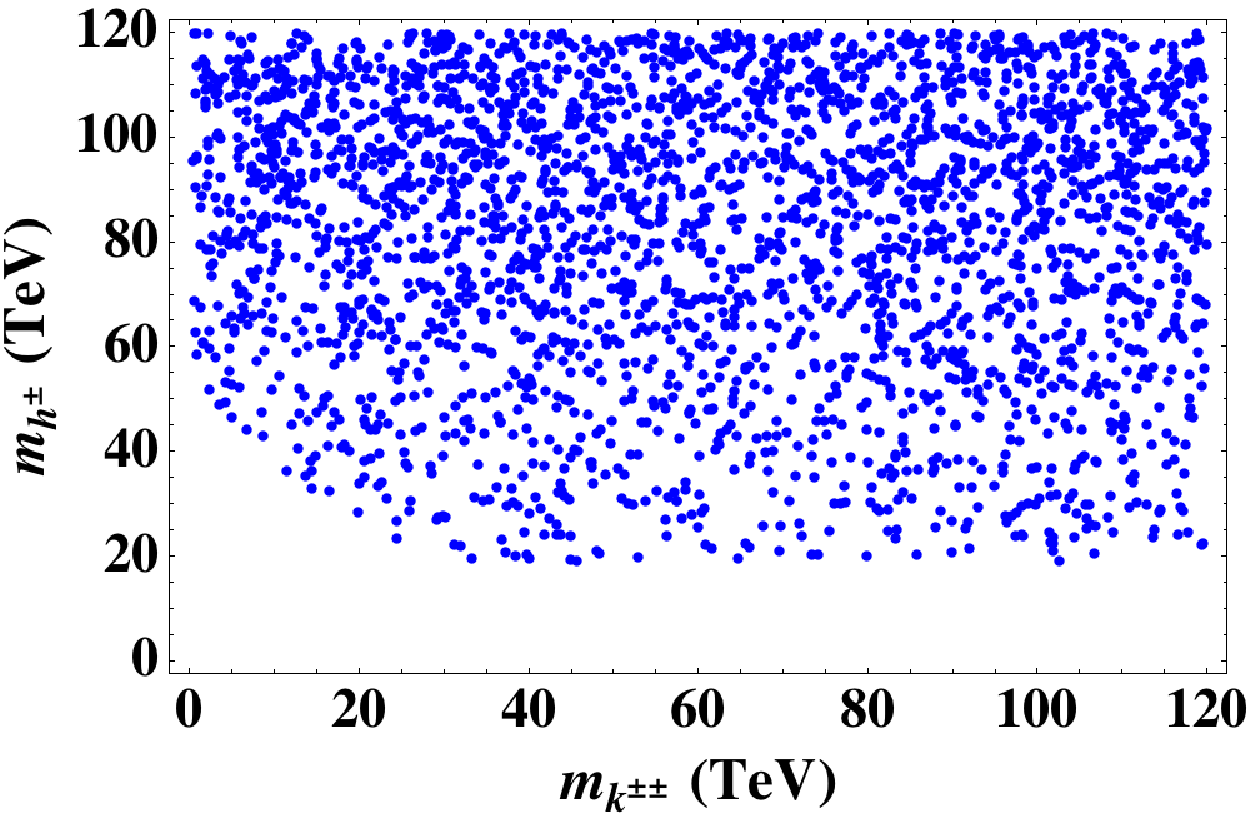}
%%%%%%%
\caption{The allowed mass {ranges} of the original {Zee-Babu} model for the {singly-charged} bosons $(h^\pm)$ and the doubly-charged boson $k^{\pm\pm}$ to satisfy all the constraints discussed in our model. These figures {tell} us that the allowed region is much wider than our allowed one for both of {the orderings}, where the left figure is the {NH} case, while the right one is the {IH} case.
Note that we examine $10^5$ points in this scanning.
}
   \label{fig:mass_ZB}
\end{center}
\end{figure}

Figure~\ref{fig:mass_ZB} shows the allowed region of {$m_{k^{\pm\pm}}$ and $m_{h^\pm}$} in the original {Zee-Babu} model in order to compare with our allowed regions, where one finds that the allowed parameter configurations are much wider than ours in both of the orderings. Here, we take the same parameter set in Eq.~(\ref{paraset}).

%%%%%%%%%%%%%%%%%%%
%%%%%%%%%%%%%%%%%%%

%%%%%%%%%%%%%%%%%%%
%%%%%%%%%%%%%%%%%%%

%%%%%%%%%%%%%%%%%%%

%%%%%%%%%%%%%%%%%%%
%%%%%%%%%%%%%%%%%%%

%%%%%%%%%%%%%%%%%%%%%%%%%%%%%%%%%%%%%%%%%%%%%%%%%%%%%%%%%%%%%%%%%%%%%%%%%%%%%%
%%%%%%%%%%%%%%%%%%%%%%%%%%%%%%%%%%%%%%%%%%%%%%%%%%%%%%%%%%%%%%%%%%%%%%%%%%%%%%
\subsection{Collider-related issues}
%%%%%%%%%%%%%%%%%%%%%%%%%%%%%%%%%%%%%%%%%%%%%%%%%%%%%%%%%%%%%%%%%%%%%%%%%%%%%%
%%%%%%%%%%%%%%%%%%%%%%%%%%%%%%%%%%%%%%%%%%%%%%%%%%%%%%%%%%%%%%%%%%%%%%%%%%%%%%

In this subsection, we discuss possible collider-related issues.
In our model, apart from the Zee-Babu model, the masses of the charged scalars are apt to be very heavy as typically above $10\ \text{TeV}$.
They are highly decoupled and it is very difficult to detect a signature of our model even in the $14\ \text{TeV}$ Large Hadron Collider (LHC) at CERN.
However, still some rooms remain in colliders.

First, we consider the constraint from Higgs search at the LHC.
Now, the SM-like Higgs boson was discovered~\cite{Aad:2012tfa,Chatrchyan:2012ufa} and the signal strengths in various channels have been measured precisely.
The latest value of the diphoton decay channel in the ATLAS experiment is announced as $\mu_{\gamma\gamma} = 1.17 \pm 0.27$ with the central value of the Higgs mass $125.4\ \text{GeV}$ based on the data taken in the $\sqrt{s} = 7\ \text{TeV}$ and $\sqrt{s} = 8\ \text{TeV}$ LHC corresponding to a total integrated luminosity of $25\ \text{fb}^{-1}$~\cite{Aad:2014eha}.
The CMS counterparts are $\mu_{\gamma\gamma} = 1.14^{+0.26}_{-0.23}$ and $124.70\ \text{GeV}$, where the integrated luminosities of the data samples are $5.1\ \text{fb}^{-1}$ at $\sqrt{s} = 7\ \text{TeV}$ and $19.7\ \text{fb}^{-1}$ at $\sqrt{s} = 8\ \text{TeV}$~\cite{Khachatryan:2014ira}.

In our model, we find three charged scalar particles, $k^{\pm\pm}, h_1^{\pm}, h_2^{\pm}$ coupling with the photon and the Higgs boson and their contributions modify the signal strength of the Higgs diphoton decay.
Besides, the CP-even physical component of the {$SU(2)_L$} doublet $\phi$ is mixed with the corresponding part of $\Sigma_0$, $\sigma$, which is never introduced in the Zee-Babu model, and consequently the factor via the mixing, $\cos{\alpha}$ in Eq.~(\ref{eq:mass_weak}), deforms the couplings with respect to the SM-like Higgs boson $h$ defined as a mass eigenstate.

With taking into account the points, we can write down the following form like in the Zee-Babu model~\cite{Ellis:1975ap,Shifman:1979eb,Carena:2012xa,Herrero-Garcia:2014hfa},
\begin{align}
\mu_{\gamma\gamma} = \frac{\Gamma(h \to \gamma\gamma)_{\text{ours}}}{\Gamma(h \to \gamma\gamma)_{\text{SM}}} =
	\left| {\cos{\alpha}} + \delta R(m_{h_{1}^{\pm}},\lambda_{\Phi h_{1}}) + \delta R(m_{h_{2}^{\pm}},\lambda_{\Phi h_{2}}) + 4 \, \delta R(m_{k^{\pm\pm}},\lambda_{\Phi k}) \right|^2,
\end{align}
where the function $\delta R(m_x, \lambda_{\Phi x})$ ($x$ standing for a type of the charged particles) is defined with the loop functions $A_{0}, A_{1/2}, A_{1}$ as
\begin{align}
\delta R(m_x, \lambda_{\Phi x}) &= \frac{\lambda_{\Phi x} v^2}{2m_{x}^2} {\cos{\alpha}} \frac{A_0(\tau_x)}{A_1(\tau_W) + \frac{4}{3} A_{1/2}(\tau_t)}, \\
%%%
A_{0}(x) &= -x + x^2 f(1/x), \\
A_{1/2}(x) &= 2x + 2x(1-x) f(1/x), \\
A_{1}(x) &= -2 -3x -3x(2-x)f(1/x).
\end{align}
Here, $\tau_i$ means $4m_i^2/m_h^2$ and the concrete form of $f(x)$ is $\arcsin^2{\sqrt{x}}$ (when $x \leq 1$) \footnote{In our case, every particle including the W boson and the top quark inside the loop fulfills the condition, $4m_i^2 > m_h^2$, being equivalent to $\tau_i \geq 1$.}.
Here in our model, all the charged scalars are decoupled and contributions from these particles are negligible as $\mu_{\gamma\gamma} \approx \cos^2{\alpha}$.
It is very easy to estimate the lower bounds on $|\cos{\alpha}|$ with $2\sigma$ confidence level from the ATLAS and the CMS experiments are $0.794$ and $0.825$, respectively, while no upper bound arises on $|\cos{\alpha}|$.
%Note that when we go down from the bounds, no parameter region can be survived.
%\footnote{
%When both of $\lambda_{\Phi k}$ and $\lambda_{\Phi h_{1,2}}$ are positive, the interference is always destructive. After we remember that $\cos{\alpha} \le 1$, we can conclude that this exclusion is valid for any heavier charged scalar in this case. When we flip either or both of the two couplings, different situations would occur.}.
This issue highly suggests that the value of $\cos{\alpha}$ should be close to one.
One possibility for realizing $\cos{\alpha} \simeq 1$ is that the mass of the additional CP-even scalar is very large (see Eq.~(\ref{eq:CP-even_mixing})).

Next, we consider the prospects in future like-sign electron linear collider.
For generating a doubly-charged scalar as an $s$-channel resonance, a like-sign linear collider is a fascinating option \footnote{
Lots of works have already been done about the physics in a like-sign linear collider, e.g., see~\cite{Rizzo:1981xx,Rizzo:1982kn,London:1987nz,Cuypers:1997qg,Raidal:1997tb,Cakir:2006pa,Rodejohann:2010jh,Rodejohann:2010bv}.}.
As discussed in~\cite{Schmidt:2014zoa}, after accumulating a total luminosity of $50\ \text{fb}^{-1}$, more than a few tens of signal events of $e^{-} e^{-} \to k^{--} \to \ell^{-} \ell^{-}$ can be expected for a doubly-charged scalar with $m_{k^{\pm\pm}} \lesssim 10\ \text{TeV}$ (even) with the center of mass energy $\sqrt{s} = 500\ \text{GeV}$ in a like-sign electron linear collider, where possibly, we cannot reconstruct the mass of the doubly-charged scalar since this particle seems to be off-shell \footnote{
Note that we can differentiate the the Zee-Babu model from {$SU(2)_L$} triplet models (including a doubly-charged scalar) by measuring the processes $e^{-} e^{-} \to \ell^{-}_\alpha \ell^{-}_\beta$ with various final states and analyzing the patterns~\cite{Schmidt:2014zoa}.
}.
Even though detailed expectations depend on the interrelation among the matrix elements of $y_R$, we can expect that our model and the Zee-Babu model are widely tested up to the parameter region with a large value in $m_{k^{\pm\pm}}$ (and also in $m_{h_{1,2}^{\pm}}$).

Now, we try to evaluate the prospects for the discovery of our model in a like-sign electron linear collider through the process $e^- e^- \to k^{--} \to e^- e^-$.
We choose the template value of $m_{k^{\pm\pm}}$ as {$20\ \text{TeV}$} (being rather close to the minimum) and assume $(y_R)_{ee} = 1$, these values are not favored for evading the bounds from lepton-flavor violation processes, but it would be realizable with fine tuning in the other elements of $y_R$. For other values, we assign {$10^{-6}$} to circumvent the bounds.
We consider the cases of $\sqrt{s} = 1,\ 3,\ 5,\ 10\ \text{TeV}$ and ignore the two {singly}-charged scalars in calculating the decay width of $k^{\pm\pm}$.
For estimating the production cross section, we implement our model with the help of {\tt FeynRules\,2.1}~\cite{Christensen:2008py,Alloul:2013bka} to generate the model file in the UFO format~\cite{Degrande:2011ua} for simulations in {\tt MadGraph5\_aMC@NLO}~\cite{Alwall:2011uj,Alwall:2014hca}.
%%%
The values of $\sigma_{e^{-} e^{-} \to k^{--} \to h_2^{-} h_2^{-}}$ are $0.0959\ \text{fb}$, $0.898\ \text{fb}$, $2.71\ \text{fb}$ and $16.9\ \text{fb}$, respectively.
%%%
Then, we can conclude that a multi-TeV like-sign electron linear collider could hold reasonable potential for exploring our model, at least in the specified choice.
After accumulating a large amount of data, the case with $(y_R)_{ee} \lesssim 1$ could be reachable.

%%%%%%%%%%%%%%%%%%%%%%%%%%%%%%%%%%%%%%%%%%%%%%%%%%%%%%%%%%%%%%%%%%%%%%%%%%%%%%
%%%%%%%%%%%%%%%%%%%%%%%%%%%%%%%%%%%%%%%%%%%%%%%%%%%%%%%%%%%%%%%%%%%%%%%%%%%%%%
%%%%%%%%%%%%%%%%%%%%%%%%%%%%%%%%%%%%%%%%%%%%%%%%%%%%%%%%%%%%%%%%%%%%%%%%%%%%%%
%%%%%%%%%%%%%%%%%%%%%%%%%%%%%%%%%%%%%%%%%%%%%%%%%%%%%%%%%%%%%%%%%%%%%%%%%%%%%%

%\section{Conclusions}
\section{Conclusions\label{sec:conclusions}}
%\textcolor{blue}
We have constructed a three-loop induced neutrino model with a global $U(1)$ symmetry, in which we naturally
accommodate a DM candidate. Taking into account for all the constraints, {namely,} the vacuum stability of these charged bosons,  all the lepton flavor violating processes, the observed neutrino masses and the {mixings} under our parameter set in Eq.~(\ref{paraset}), we have obtained allowed regions at  {$10\ \text{TeV} \lesssim {m_{k^{\pm\pm}}} \lesssim 100\ \text{TeV}$, $10\ \text{TeV} \lesssim {m_{h_{1}^\pm}} \lesssim 100\ \text{TeV}$, and $20\ \text{TeV} \lesssim {m_{h_{2}^\pm}} \lesssim 100\ \text{TeV}$} for the {NH} case,
and {$10\ \text{TeV} \lesssim {m_{k^{\pm\pm}}} \lesssim 170\ \text{TeV}$, $20\ \text{TeV} \lesssim {m_{h_{1}^\pm}} \lesssim 170\ \text{TeV}$, and $20\ \text{TeV} \lesssim {m_{h_{2}^\pm}} \lesssim 150\ \text{TeV}$} for the {IH} case, {respectively}.
We have found that the {NH} case is prone to have wider allowed region.

%%%
%On the other hand, the DM mass is expected not to be
%so large for {both of the neutrino mass orderings} that is ${\cal O}$(100) TeV to maintain the stability of these couplings, unless the the charged boson masses become much heavier or take the constraint of the vacuum stability more loosen. %in Eqs.~(\ref{cdt-lam1})-(\ref{cdt-vs}).
%In this sense, our DM can naturally explain the observed relic density  and the direct detection searches which typically lies on the ${\cal O}$(100) GeV mass scale.

In our model, the masses of the charged scalars tend to be very large as around $\mathcal{O}(10)\ \text{TeV}$ because of the required large trilinear couplings among these particles leading to the menace to the {positivity} of scalar quartic couplings at the one-loop level.
After considering the stability of the DM, its mass is naturally bounded from above.
In this sense, our DM candidate can naturally explain the observed relic density and the direct detection searches which typically lies on the ${\cal O}$(100) GeV mass scale.

%%%
We have discussed possible collider-related issues, in which, the masses of the charged scalars tend to be very heavy as above $10\ \text{TeV}$ that is apart from the Zee-Babu model.
Hence, they are highly decoupled from the SM particles and it is very difficult to detect a signature of our model even in the $14\ \text{TeV}$ LHC.
However, we expect that a future like-sign electron linear collider could detect such heavy {particles such as $k^{\pm\pm}$ (and also $h_{1,2}^{\pm}$, probably)}.

%\newpage
%%%%%%%%%%%%%%%%%%%%%%%%%%%%%%%%%%%
\vspace{0.3cm}
%\hspace{0.2cm} {\bf Acknowledgments}
%\section*{Acknowledgments}:
%\vspace{0.5cm}
\section*{Acknowledgments}

H.O. thanks to Prof.~Seungwon Baek, Dr.~Takashi Toma, and Dr.~Kei {Yagyu} for fruitful discussions.
K.N. is grateful to Prof.~Shinya Kanemura and Prof.~Tetsuo Shindou for useful communications.
This work is supported in part by NRF Research Grant 2012R1A2A1A01006053 (H.H.), No. 2009-0083526 (Y.O.) of the Republic of Korea.

%%%%%%%%%%%%%%%%%%%%%%%%%%%%%%%%%%%

\end{document}